# An Efficient Indexing and Searching Technique for Information Retrieval for Urdu Language

Mudassar[1], Shoaib[1] and Kalsoom[2]

[1]Department of CS & E, University of Engineering and Technology, Lahore, Pakistan

[2] National Library of Engineering Sciences, UET, Lahore, Pakistan

**ABSTRACT:** Indexing techniques are used to improve retrieval of data in response to certain search condition. Inverted files are mostly used for creating indexes. This paper proposes indexing technique for Urdu language. Language processing step in Index creation is different for a particular language. We discuss index creation steps specifically for Urdu language. We explore morphological rules for Urdu language and implement these rules to create Urdu stemmer. We implement our proposed technique with different implementations and compare results. We suggest that indexes should be created without stop words and also index file should be an order index file.

**Keywords:** Index technique, Inverted files, Urdu Language, Urdu Stemmer, Morphological rules

## INTRODUCTION

Indexing techniques (Ricardo and Berthier, 2004) are used for fast retrieval of data in search operations. There are many indexing techniques exists and each indexing technique is appropriate for a specific situation. Inverted file (Chouvalit and Veera, 2007) (Fidel and Angel, 2002) is the most common indexing technique and used for common operations. But this indexing technique is appropriate for English language and if we want to make a search system for other than English language such as Urdu language, then this indexing technique is not suitable. So for searching contents in Urdu language, we have to create indexing technique specific for Urdu language (Lee-Feng and Hsiao-Tieh, 1996).

"Information retrieval (IR) is the science of searching for information in documents, searching for documents themselves, searching for metadata which describe documents, or searching within databases, whether relational stand-alone databases or hyper textually-networked databases such as the World Wide Web".(Christopher *et al*, 2008). Information retrieval (IR) objective is to store, organize, represent information contents and provide access to that information contents.

Language processing is major step in index creation. This step has sub steps (Stop words removal, normalization and stemming). We identified stop words list for Urdu language. Normalization is application specific functionality. In stemming process, we have to create morphological rules to create a stemmer. We create Urdu Stemmer by identifying morphological rules for Urdu language. We implement our proposed indexing technique and create index file. Different implementations are used and we compare our results. We suggest that we have to remove stop words from index file so that retrieval of records is fast. Also implement stemming rules so that more relevant results are retrieved. Also we suggest that an ordered indexing file is needed for fast retrieval of records.

## MATERIALS AND METHODS

### *Indexing*

Searching algorithms are used in many applications but there are some limitations of these algorithms. When database is huge or very large text document, the retrieval becomes slow down. The solution to this problem is to build data structure (index) for fast searching. There are many indexing techniques but most familiar techniques are Inverted files, Suffix arrays, Signature files.

### Inverted Files

An inverted file is used to store contents as mapping of different documents. Mapping is done with two key elements **Vocabulary** and **Occurrences.**

### *Index Creation*

Index creation is an important task in information retrieval. Index creation is a process which has different steps.

### Steps for Index Creation

Major steps in creating index are (Christopher *et al,* 2008)

1) Document collection that will be used for index

2) Text tokenization

3) Language processing for tokens

4) Index those documents in which each term lies

### Document Collection

For creating document collection, we have to create digital documents. Digital documents are files which consist of bytes in a computer system. For processing first step is to convert these bytes into characters. In case of English, this is simple process. But if a document is encoded in Unicode or UTF-8, it will be a complex step. For details, please refer to (Christopher *et al*, 2008).

### Tokenization

Tokenization means splitting of given document into pieces called tokens.

For example,

Input: Ali is a brave boy.

Output: | Ali | is | a | brave | boy |

### Language Processing

### *Stop Words*

In any language, some very common words are present which have little contribution for processing of query, so we eliminate those words to get efficient result. These common words are called stop words. For further details please refer to (Christopher *et al*, 2008). In English, examples are (a, an, the, prepositions (by, at, for), and, why, etc). In Urdu examples are (کیوں، ہے،یہ، کیسے).

### *Normalization*

We can make efficient index by using normalization process. In this process, we make equivalent classes of same term. For example, word UK. We also want to make this word searchable when querying U.K. The standard way is to create implicit equivalence classes. For example, UK and U.K are both mapped to UK.

## *Stemming*

"Stemming usually refers to crude heuristic process that chops off the ends of words in the hope of achieving this goal correctly most of the time and often includes the removal of derivational affixes"(Christopher *et al*, 2008).In simple words, in stemming our goal is to reduce inflected or derived words to their stem or base word.

There are different forms of words due to grammatical reasons. For example in English word, "transfer" has different forms like transferred, pre transfer, post transfer, transfers.

Some Urdu words have two or more plural forms like word duurra (period,دورہ) has different plural forms like:-

Duuree (periods, دورے)

Duuroon (periods, دوروں)

Adwaar (periods, **ادوار**)

There is almost no work present in literature for index creation in Urdu Language (Kashif, 2007).

There are different stemming algorithms for English language. Most popular stemming algorithm is Porter Stemmer (Porter,1980). Other stemming algorithms are Lovins Stemmer (Julie, 1968), Paice/Husk stemmer, Dawson Stemmer and Krovetz algorithm.

## *Indexing Technique for Urdu Language*

As we discussed earlier, there is hardly any work presented for Urdu language in the field of IR (Especially in Index Creation process) which we use to create our proposed indexing technique. First step is collection of documents. These documents are source from which we search against user queries. These documents are converted in electronic format. Next step is tokenization of these documents. We tokenize documents on basis of white space means a token is a word in a document. Next step is removing stop words from our tokens. Next we normalize our tokens. In next step we use these normalized tokens to create stemmer. Then these normalized and stemmed tokens are used to create inverted index file. This file is used to make our proposed system more efficient because due to use of this file, system provides more efficient and relevant results against user query. The architecture of proposed indexing technique (Mudassar, 2010) is shown in Figure1.

**Figure- 1: Architecture of Urdu Information Retrieval System**

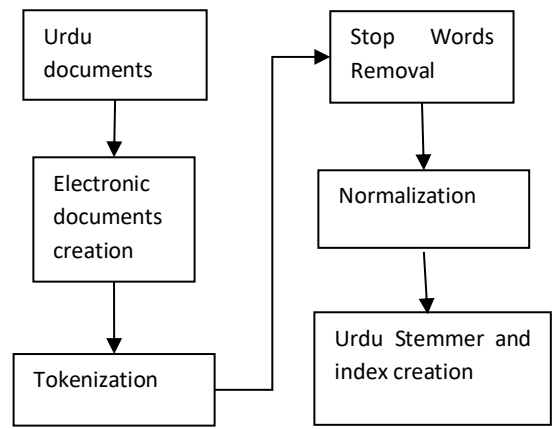

## *Step 1: Document Collection*

First step is collection of documents. We used Unicode UTF-8 encoding scheme for making electronic documents that are used for processing of our proposed indexing technique.

### Step 2: Tokenization

In this step we generate tokens from electronic documents. For instance, one record in temporary storage is

"وہ لاہور آیا"

Then we generate three tokens from above record.

1) وہ
2) لاہور
3) آیا

### Step 3: Stop Words Removal

As we discussed earlier stop words have a little use in IR system, so we eliminate these words. We used stop words table to reduce the size of index file. For example, if user enters a query

"وہ لاہور آیا"

We can eliminate "وہ" word from query using stop words list.

### Step 4: Normalization

This step is application specific. Means we have different process in this step for different applications.

### Step 5: Stemmer and index Construction

We have to define rules in order to create stemmer. Our objective is to create stemmer for Urdu language, so we have to define rules for Urdu language. These rules should be entirely different from English and other Native Languages because these grammatical rules vary from one language to other language. We use grammatical rules of Urdu present in Urdu grammar literature and also use morphological analysis of Urdu grammar to come up with rules that help in construction of Stemmer for Urdu language. In this section, first we present rules for stemmer (Sara, 2004).

There are two main groups of verbs with respect to their ending alphabet.

1. Verbs that end with consonant alphabet (Alphabet except ا،و،ی،ے).
2. Verbs that end with alphabet ا،و،ی،ے. We will call it vowel group.

From second group, there are further subgroups:

I. Verbs that end with alphabet او
II. Verbs that end with alphabet ی
III. Verbs that end with alphabet ے

**Step 5.1**

This step is related to plurals in Urdu.

This step has further five sub steps

**Step 5.1.1:** *Tokens are sequentially scanned and check their last character. If it is consonant then append ­وں at the end and* search for its location and save in index list.

**Step 5.1.2:** *Tokens are sequentially scanned and check their last character. If it is ا then delete ا and append ےوں at the end and* search for its location and save in index list.

**Step 5.1.3:** *Tokens are sequentially scanned and check their last character. If it is ی then append یاں، وں at the end and* search for its location and save in index list.

**Step 5.1.4:** *Tokens are sequentially scanned and check their last character. If it is ہ then delete ہ and append ےوں at the end and* search for its location and save in index list.

**Step 5.2**

In this step, we cover inflected forms of verb with respect to tenses in Urdu.

For example

لکھنا -----> لکھائ

-----> لکھتا

-----> لکھتی

Here لکھنا is definitive verb and لکھائ,لکھتا, لکھتیare inflected verbs. Here root verb is لکھ and it is obtained by removing (نا) from definitive verb. This word (نا) is called auxiliary word. We identify some auxiliary words which are (نا,نے, نی, ا, ی, ے, تا, تی, تیں, یں, و, یے )

This step has further four sub steps.

**Step 5.2.1:** In this step, we identify rules for consonant verbs (verbs except ا،و،ی،ے). *Tokens are sequentially scanned and check their last two characters. If it is an auxiliary word then delete auxiliary word and search for its location and save in index list.*

**Step 5.2.2:** In this step, we identify rules for ا،و. *Tokens are sequentially scanned and check their last two characters. If it is an auxiliary word then delete auxiliary word and search for its location and save in index list.*

**Step 5.2.3:** In this step, we identify rules for ی. *Tokens are sequentially scanned and check their last two characters. If it is an auxiliary word then delete auxiliary word and search for its location and save in index list.*

**Step 5.2.4:** In this step, we identify rules for ے. *Tokens are sequentially scanned and check their last two characters. If it is an auxiliary word then delete auxiliary word and search for its location and save in index list.*

## RESULTS AND DISCUSSION

Below are some results which are obtained from our computational model.

1) We created indexing file for two times. First time, we create file with stop words. In that case size of file is 461 kilo bytes (KB). Second time, we create file without stop words, then it size is 453 kilo bytes (KB).

2) We observe results for three different implementations in order to compare efficiency of indexing technique in respect of time execution. First implementation is "with stop words", means stop words are present in indexing file. Second implementation is "without stop words and unordered". This means indexing file has no stop words but file is unordered. And third implementation is "without stop words and order". This means that indexing file has no stop word and indexing file is ordered. Table 1 shows execution time of some results of these three implementations. These results are graphically represented in Figure2.

**Table- 1: Time Execution Comparison**

| Word | With stop words (micro seconds) | Without stop words and unordered (micro seconds) | Without stop words and order (micro seconds) |
|---|---|---|---|
| گناہ | 0.43 | 0.42 | 0.21 |
| نماز | 0.70 | 0.52 | 0.27 |
| جنت | 0.45 | 0.47 | 0.27 |
| رب | 1.38 | 0.79 | 0.52 |

**Figure- 2: Time Execution Comparison**

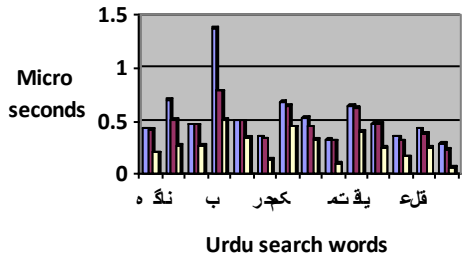
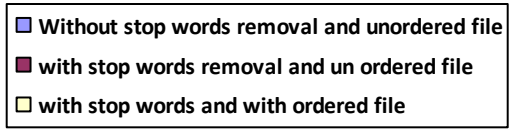

3) We also observe results without implementing stemming rules and implementing with stemming rules. Table 2 shows records of some results of these two implementations. These are graphically represented in Figure 3.

**Table- 2: Comparison of retrieved results**

| Word | Records without stemming rules | Records with stemming rules |
|---|---|---|
| جنت | 41 | 59 |
| رات | 42 | 48 |
| آسمان | 128 | 389 |
| کافر | 112 | 279 |
| کتاب | 158 | 193 |

**Figure- 3: Comparison of retrieved results**

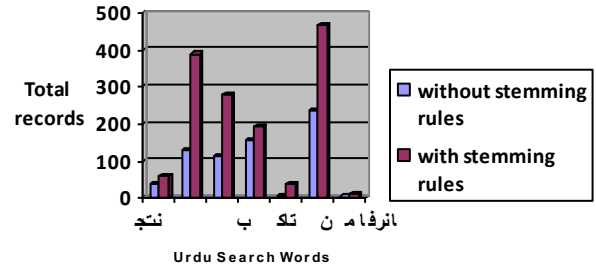

We see that in above section, size of indexing file without stop words is less than the indexing file without stop words. Also stop words have a little or no use in searching. Also we see that execution time of searching is lower when we use indexing file without stop words. So we recommend that index file should be used without stop words.

We can observe that execution time of searching "without stop words and order" is less than "without stop words and order" implementation. So we can infer that if we implement ordering in indexing file, than it is more efficient then if we do not implement any order in indexing file. So we recommend that, ordering should be performed in creating indexing file. One thing, we should note is that ordering is done on basis of how we want to use that indexing file. For example, in our computational model we want to use that indexing file for searching of Urdu contents which are only related to Urdu alphabets. So we order that indexing file purely on basis of Urdu alphabets. If someone wants to implement indexing file for searching of hotels on basis of each country, he should order that indexing file on basis of country names.

Table 5.2, we show records for each result. These are total records retrieved for each query word. For example there are 41 records retrieved against word "جنت" for Column1 ("Without stemming rules") and 59 records for Column2 ("with stemming rules. In Column1, we do not implement stemming rules, and Column2, we implement stemming rules. We implement stemming rules, if we want to allow users for expanded search. This means that if user search for word "work", we can allow user to search also for related words i.e. "works, worked, worker ,etc". So, we can infer that, for expanded search, we have to implement stemming rules in our indexing technique.

Keeping in view of expanded search, we can make our indexing technique more efficient by introducing multilevel indexing for stem words. For example we have a word "جنت". We can create multilevel index for this for other words related to it .i.e " جنتیں,جنتوں ".

## *Conclusion & Recommendations*

We describe in detail our proposed indexing technique. We suggest that our proposed indexing technique will be helpful for fast retrieval of records in searching of Urdu contents.

We explore stemming rules only for verbs. But we do not explore some complex rules such as special cases of plurals. Also we do not explore complex rules of verb which mostly relate to adverbs. One can carry from this point and can explore more rules. Also work can be done in exploring nouns of Urdu language that their behavior is how much deviate from verbs.